# Tuning the properties of magnetic CdMnTe quantum dots

S. Mackowski*, H. E. Jackson, and L. M. Smith

*Department of Physics, University of Cincinnati, Cincinnati, OH 45221-0011*

J. Kossut and G. Karczewski

*Institute of Physics, Polish Academy of Science, Warsaw, Poland*

W. Heiss

*Institut für Halbleiter- und Festkörperphysik, Johannes Kepler Universität Linz, Austria*

*We show that CdMnTe self-assembled quantum dots (QDs) can be formed by depositing a submonolayer of Mn ions over a ZnTe surface prior to deposition of the CdTe dot layer. Single dot emission lines and strongly polarized QD photoluminescence (PL) in an applied magnetic field confirm the presence of Mn in individual QDs. The width of PL lines of the single CdMnTe dots is 3 meV due to magnetic moment fluctuations (MMFs) of the Mn ions. After rapid thermal annealing (RTA), the emission lines of individual magnetic QDs narrow significantly to 0.25 meV showing that effect of MMFs is strongly reduced most probably due to an increase in the average QD size. These results suggest a way to tune the spin properties of magnetic QDs.*

*Author to whom the correspondence should be addressed: electronic mail: seb@physics.uc.edu

Quantum dots (QDs) containing magnetic ions hold significant promise for performing logic operations utilizing electron spins [1]. Several ways of fabricating magnetic QDs have been reported recently, including self-assembly [2,3,4], annealing of specially designed structures [5], or synthesis of nanocrystals [6]. Self-assembled magnetic QDs are particularly attractive for this purpose, because of their high structural quality and excellent optical properties [3]. Three-dimensional spatial confinement in these QDs is evidenced by the observation of single dot emission lines; these lines are much broader (~ 3-4 meV) [7,8] than usually observed for non-magnetic QDs (~0.1 meV). This significant increase in the linewidth is caused by magnetic moment fluctuations (MMFs) within a single magnetic QD [9], which can be suppressed by applying a strong magnetic field [9]. Alternatively, the MMFs that broaden single DMS QD emission lines are expected to be significantly smaller for larger dots [9], probably due to the decreased sensitivity of the QD emission energy to the configuration of Mn ions in a larger QD. The suppression of MMFs may then result in narrow emission lines of a single magnetic QD even at B=0 T. In addition, magneto-optical measurements demonstrate large magnetic field-induced Zeeman splitting of electronic levels in CdMnSe QDs resulting in strong polarization of the emission lines [3]. These effects, characteristic of diluted magnetic semiconductors (DMS), are a consequence of the strong exchange interaction between the carriers and the d-electrons localized on Mn ions [10]. Theoretical studies suggest that the strength of this interaction depends strongly on the carrier confinement [9,11] and that the spin effects should be significantly enhanced for DMS-based QDs [11,12].

In this work, we show that CdMnTe QDs are formed when a ZnTe buffer layer is covered with Mn before depositing the CdTe QD layer. Both predominantly $\sigma^+$ – polarized PL in magnetic field and observation of single dot emission lines in μ-PL spectra prove the presence of Mn in QDs. After RTA we observe a dramatic narrowing of the single CdMnTe dot emission lines from 3 meV for the as-grown QDs to 0.25 meV for the annealed QDs, which we ascribe to the suppression of MMFs-induced PL broadening for larger DMS QDs.

The samples (see the inset in Fig. 1) were grown by molecular beam epitaxy on (100)-oriented GaAs substrates. Prior to deposition of the CdTe dot layer, the ZnTe surface was partially covered by Mn ions by opening the Mn shutter for several (from 0 to 9) seconds. The QDs were then grown by then depositing 4 monolayers of CdTe at a temperature of 320C. In order to change the average QD size, the samples were annealed in an Ar atmosphere for 15 seconds at temperatures from $350^0$C to $500^0$C.

The optical properties of the QDs were examined by PL, µ-PL and time-resolved PL. Continuous-wave PL experiments were performed at a temperature of T=4.2 K in magnetic field up to B=4 T applied in the Faraday configuration. A Babinet-Soleil compensator and Glan-Thomson linear polarizer were used to analyze the circular polarization of the emission. The laser was focused on the sample with a diameter of 30 µm or 1 µm by a lens or a microscope objective, respectively. The emission was dispersed by a DILOR triple monochromator and detected by a CCD detector. The recombination dynamics at T=2 K was studied by time-correlated single photon counting. A frequency-doubled Ti: Sapphire laser giving 6 ps long pulses at a wavelength of 370 nm was used for excitation. The signal was dispersed by a 0.25 m monochromator and detected by a microchannel plate photomultiplier tube.

In Fig. 1 we present PL spectra of the QD samples grown with different Mn exposure times. While the spectrum of non-magnetic CdTe QDs features a single inhomogeneously broadened line, the PL of the Mn-doped QDs reveals a more complicated structure. The results of µ-PL and time-resolved PL experiments [8] show that the shaded regions in Fig. 1 originate from the recombination in CdMnTe QDs. When probing a micrometer-sized area of the sample, we observe that only this emission splits into a series of distinguishable lines ascribable to single dot PL [8], as typically observed for other QDs [13]. In addition, the decay times measured for the shaded emissions are 300 ps, which is a typical value for other II-VI QDs [13]. In contrast, the shape of other PL bands does not change when reducing the laser spot size, so that we do not associate them with QD recombination. Moreover, the decay times measured for these lines are longer than 12 ns. We therefore assign this long-lived emission to an intra-Mn transition [14].

The presence of the intra-Mn transition strongly reduces the PL intensity of the CdMnTe QDs. As seen in Fig. 1, the intensity ratio between these two emissions increases significantly with the time of Mn exposure. For the sample grown with 9 second long Mn exposure, the intra-Mn transition completely dominates the spectrum. This strong reduction of the QDs PL intensity could be caused by impairment of the QDs' structural quality due to high Mn content, as reported for CdMnSe QDs grown in a similar way [2]. However, this does not seem to be the case since when exciting QDs *below* the intra-Mn transition, the intensity of the QD emission is comparable for all three samples [15]. We conclude that incorporation of Mn into CdTe QDs does not affect significantly the structural quality of QDs, although the presence of the intra-Mn transition strongly suppresses the CdMnTe QDs PL.

To demonstrate that the excitons localized by QDs interact with Mn ions, we have measured polarization – resolved PL in a magnetic field. In Fig. 2 we show the spectra obtained for CdMnTe QDs at B=0 T (Fig. 2a) and B=4 T (Fig. 2b) under *linearly polarized* excitation above the ZnTe barrier. In the inset to Fig. 2, we plot the polarization P, defined as $(I^+-I^-)/(I^++I^-)$ obtained for both non-magnetic CdTe QDs (triangles) and magnetic CdMnTe QDs (circles) as a function of magnetic field. In this expression $I^+$ ($I^-$) correspond to intensity of $\sigma^+$ ($\sigma^-$) emission.

For CdTe QDs, P is essentially zero for all magnetic fields. This indicates that for above the barrier excitation, the excitonic levels in CdTe QDs, although split by ~0.6 meV at B=4T [8], are equally populated.

In contrast, for DMS QDs (circles) P is *positive* and increases with magnetic field. The positive value of P indicates the *positive* value of effective g-factor for DMS QDs, in contrast to CdTe non-magnetic QDs [8]. This effect, characteristic of DMS systems [10], is due to the exchange interaction between carriers and Mn ions. Since the excitons are strongly localized by QDs, they are able to form exciton magnetic polarons (MP) by spontaneously aligning Mn ions [16]. The formation of MP then lowers significantly the energy of the system [16]. At B=0 T, where there is no preferential direction of MP alignment, we observe identical intensities of both σ+ and σ− polarized PL of the CdMnTe QDs (see Fig. 2a). On the other hand, at B=4 T the MP tend to align according to magnetic field direction. This preference leads to the observed in Fig. 2b predominant σ+ - polarized emission of CdMnTe QDs.

The results of PL in magnetic field demonstrate the presence of the exchange interaction between the Mn ions and localized excitons. In order to study the influence of spatial confinement on the properties of these QDs we annealed the sample with the lowest Mn-concentration (shortest Mn exposure time) (see Fig.1). Recent studies [17] show that upon RTA the emission of CdTe QDs shifts towards higher energies and the PL inhomogeneous linewidth narrows. This has been interpreted as resulting from an RTA-induced increase of QD size.

In Fig. 3, we compare μ-PL spectra of CdMnTe QDs annealed at $T=420^0 C$ and $T=500^0 C$ with the as-grown sample. The qualitative behavior of the CdMnTe QDs is similar to that observed for CdTe QDs [17]. After RTA the CdMnTe QD emission shifts toward higher energies and the PL inhomogeneous linewidth narrows. The reduced linewidth indicates a narrower distribution of QD energies in the ensemble and an increase in the average size of the QDs. In addition, the recombination time of excitons confined to the CdMnTe QDs (see inset of Fig. 3) becomes shorter with increasing RTA temperature. The energy blue-shift and the decrease of the decay time suggest, in a close analogy to results of Ref. 17, that after annealing the QDs potentials are shallower due to interdiffusion. These observations show that RTA strongly affects the exciton confinement in the CdMnTe QDs.

The µ-PL spectra of single QDs presented in Fig. 4 show a striking modification of the magnetic CdMnTe QDs after annealing. Single emission lines in the as-grown magnetic dots are broad with a linewidth of 3 meV due to MMFs in a single QD [7,9]. In contrast, for annealed CdMnTe QDs much narrower lines appear in the µ-PL spectra. We observe lines as narrow as 0.25 meV, which, *although still broader* (~4x) than for CdTe QDs, are comparable to the emission linewidths of single non-magnetic QDs [13]. We note that for as-grown CdMnTe QDs such narrow emission lines are *never* observed. We attribute these narrow lines to CdMnTe QDs with RTA-induced larger lateral sizes. Indeed, theoretical studies indicate that the effect of the MMFs can be strongly reduced for larger CdMnTe dots [9]. One cannot, however, exclude the possibility that upon annealing the Mn ions escape from the QDs, which could result in similar behavior. A very recent observation by X. Tang *et al.* of narrow emission for QDs with nominally one Mn ion per QD would appear to support such a possibility [4]. However, we suggest that it is the larger size of the dots which dominates these changes (see Fig. 4) as the Mn ions are buried within the sample and rapid escape of Mn from the dots during RTA is rather unlikely. In particular, still significant broadening of the single emission lines (~0.25 meV) demonstrates the presence of Mn in annealed CdMnTe QDs.

In conclusion, we have used polarized PL as a function of magnetic field to demonstrate both the zero-dimensional character and the strong carrier – Mn exchange interaction in self-assembled CdMnTe QDs. We show that the linewidth of single CdMnTe QD emission lines that is due to MMFs, can be significantly reduced by an RTA-induced increase of QD size, and/or lower Mn content in a dot. RTA-induced changes of magnetic properties of CdMnTe QDs open interesting opportunities to study the influence of varied electronic confinement on spin-related phenomena in DMS QDs.

Acknowledgements

The work was supported by NSF grants nr 9975655 and 0071797 (United States), ÖAD project Nr. 05/2001 (Austria) and the projects SPINOSA and by PBZ-KBN-044/P03/2001 (Poland).

Figure 1. PL spectra of CdMnTe QDs grown with different Mn exposure time. Shaded areas represent emission of CdMnTe QDs. Inset: schematic of the sample structure.

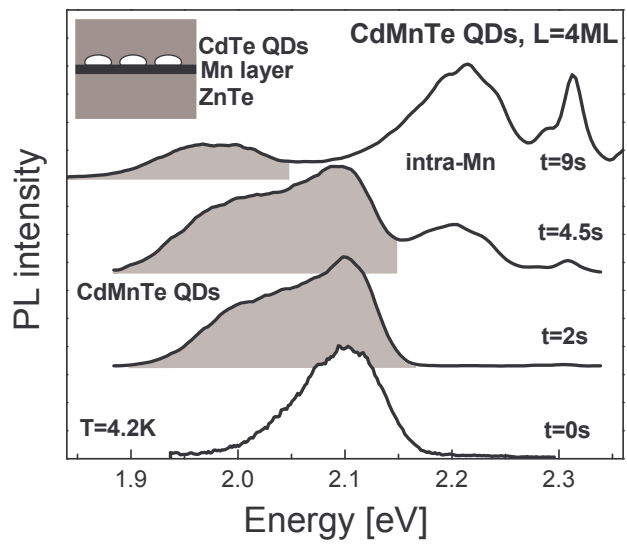

Figure 2. PL spectra for both circular polarizations at (a) B=0 T and (b) B=4 T for CdMnTe QDs grown with 2 second long Mn exposure. Inset: polarization P measured for CdTe QDs (triangles) and CdMnTe QDs (diamonds) as a function of magnetic field.

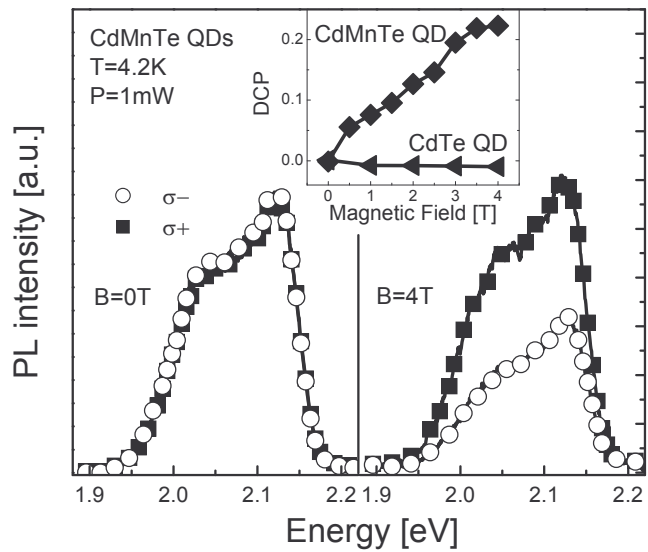

Figure 3. PL spectra for CdMnTe QDs annealed at T=420°C and T=500°C compared to the as-grown sample. Inset: temporal behavior of the PL intensity of as grown (circles) and annealed (squares) CdMnTe QD samples.

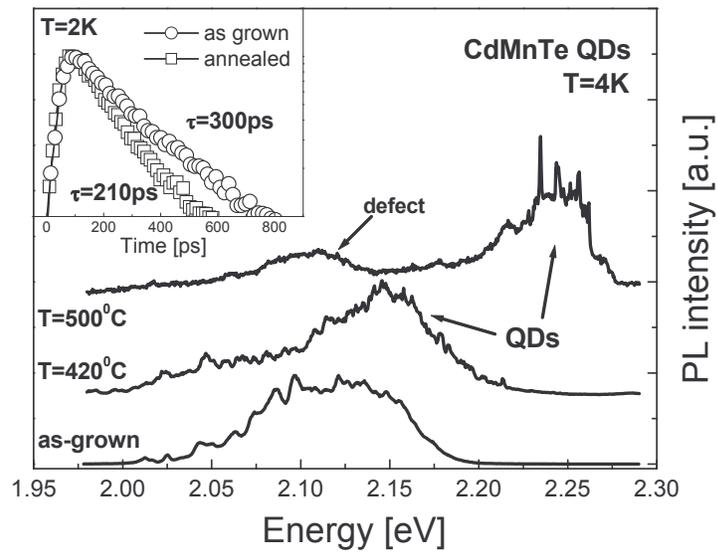

Figure 4. Single dot emissions obtained at T=4 K for as-grown and annealed CdMnTe QDs. The emission of a single CdTe QD is also shown for comparison.

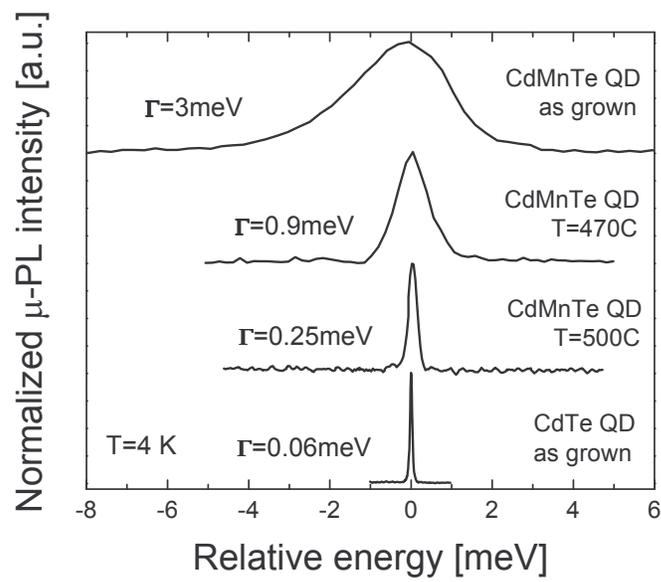